\documentclass[epj]{svjour}
\usepackage{graphics}
\begin{document}
\title{Kaon Absorption from Kaonic Atoms and Formation Spectra of Kaonic Nuclei}
\subtitle{}
\author{Junko Yamagata \and Satoru Hirenzaki
}                     
\offprints{}          
\institute{Department of Physics, Nara Women's University, Nara 630-8506, Japan}
\date{Received: date / Revised version: date}
%
\abstract{
We considered the kaon absorption from atomic states into nucleus.
We found that the nuclear density probed by the atomic kaon significantly depends on the kaon orbit.
Then, we reexamined the meanings of the observed strengths of one--body and two--body kaon absorption, and investigated the effects to the formation spectra of kaon bound states by in--flight ($K^-,p$) reactions.
As a natural consequence, if the atomic kaon probes the smaller nuclear density, the ratio of the two--body absorption at nuclear center is larger than the observed value,  and the depth of the imaginary potential is deeper even at smaller kaon energies as in kaonic nuclear states because of the large phase space for the two--body processes.
} 
\maketitle
\section{\label{sec:intro}Introduction}
Kaonic atoms and kaonic nuclei are known to 
carry important information concerning the $K^-$--nucleon interaction in nuclear medium.  This information is very important to know the kaon properties at finite density. In recent years, there have been many researches in the studies of kaonic nuclear states, which are kaon--nucleus bound systems by the strong interaction inside the nucleus. Experimental studies of the kaonic nuclear states using in--flight (${\bar K},N$) reactions were proposed and performed by Kishimoto and his collaborators \cite{Kishimoto99,Kishimoto03}. And the first theoretical results of the energy spectra of the in--flight (${\bar K},N$) reaction were obtained in Ref.~\cite{gata05} and later in Ref.~\cite{gata06} with the Green function method, where we have shown the difficulties to obtain clear signals for kaonic nuclear states formation experimentally.
Indications of $K^-pp$ bound state were reported by the FINUDA experiment \cite{finuda}. There are also theoretical studies of the structure and formation of kaonic nuclear states related to these experimental activities \cite{Akaishi02}. It should be noted that these theoretical studies predict the possible existence of ultra--high density states in kaonic nuclear systems \cite{Akaishi02,dote04}. The critical analyses of the latest data were also reported by Oset and Toki and their collaborators~\cite{oset06,magas06}. 
In our previous papers~\cite{gata05,gata06}, we have made clear that the signals of the kaonic nuclear state formation in the (${\bar K}, N$) reactions are expected to be very small. These could be complemental results to those of Oset and Toki who claimed that the origins of the structure in the experimental spectra can be explained by the well--known processes \cite{oset06,magas06}.
The detailed analysis was also performed to understand fully the FINUDA data~\cite{finuda06}.
We also mention here that some indications of existence of the NARROW kaonic nuclear states reported in Refs.~\cite{Iwasaki03} were reanalyzed by the authors leading to the different conclusions, recently~\cite{iwasaki06}.

The $K^-$--nucleus interaction has been studied for a long time based on the kaonic atom data obtained by the X--ray spectroscopy.
In Ref.~\cite{batty97}, the phenomenological study of kaonic atoms are performed comprehensively, where the density--dependent potentials are considered for $\chi ^2$ fitting to take into account possible non--linear effects which could be due to $\Lambda(1405)$ resonance.
There are also $K^-$--nucleus theoretical interactions based on the SU(3) chiral Lagrangian~\cite{waa,lut,ram,ciep01,tolo01,tolo02,schaffner00}.
These theoretical potentials are shown to have the ability to reproduce the kaonic atom data reasonably well~\cite{hirenzaki00,baca00}.
The kaonic nuclear states were also studied using these interactions and shown to have large decay widths of the order of several tens of MeV.~\cite{hirenzaki00,friedman99,friedman99_2}

One of the most important physical effect to realize the NARROW kaonic nuclear states assumed in Refs.\cite{Kishimoto99,Akaishi02,dote04} is the suppression of the imaginary potential for deeply bound kaonic nuclear states due to the threshold effects of the decay processes.
Actually the one--body process ${\bar K}N\rightarrow \pi \Sigma$ does not occur if the total energy is smaller than the threshold energy $E=m_{\bar K}+m_N-101~{\rm MeV}$, and as for the two--body process ${\bar K}NN\rightarrow \Sigma N$, the threshold energy is $E=m_{\bar K}+2m_N-239~{\rm MeV}$, where we have also used the average masses of their charged states for $\pi,~N,~\Sigma$ hadrons.
Thus, if the ${\bar K}$ has much smaller energies in the nucleus than that in vacuum, these decay channels could be suppressed by the threshold effects.
In the bound states of some of the kaonic nuclear states predicted in Refs.~\cite{Akaishi02,dote04}, the one--body decay processes could be closed.
However, the two--body processes are still open at those energies and provide the imaginary part of the kaon--nucleus potential, which can be expected to have larger effects than one--body processes for higher nuclear densities such as predicted in Refs.~\cite{Akaishi02,dote04}.
Thus, we think that it is very important to know correctly the strength of one--body and two--body decay channels for the studies of the kaonic nuclear states.

The purpose of this paper is to study the kaon absorption in nuclei from atomic orbits and to reexamine the meanings of the experimental data.
We also like to evaluate the importance of one--body (${\bar K}N\rightarrow \pi Y$) and two--body (${\bar K}NN\rightarrow YN$) decay channels of kaonic nucleus from the experimental data in a phenomenological way for a specific nuclei ($^{12}$C).
The comprehensive analysis for all existing data like performed in Ref.~\cite{batty97} is beyond the scope of this paper.
The evaluation considered in this paper, however, is significantly important to know the density and energy dependence of the imaginary part of the ${\bar K}$--nucleus optical potential. Since the predictions of the existence of the narrow kaonic nuclear states are essentially based on the imaginary potential suppression due to the threshold effects of the decay channels, we need to know the strength of each decay channel which has different energy dependence, and to know the correct threshold effects. In addition, we need to know the density dependence of the optical potential to evaluate the widths of the kaonic nuclear states in the possible high--density matter predicted in Refs. ~\cite{Akaishi02,dote04}, where the multi--body processes should be more important for kaonic nuclear decay processes and widths.
Then, we also like to show the effects to the ($K^-,p$) spectra reported in Ref.~\cite{gata06}.

In Sec.~\ref{sec:2}, we show the density distributions of kaonic atoms to know the effective nuclear densities probed by the kaon absorptions in the stopped ${\bar K}$ experiments.
In Sec.~\ref{sec:3}, we evaluate the relative strength of one--body and two--body absorptive processes and obtain the optical potentials in a phenomenological way.
In Sec.~\ref{sec:4}, we show the expected (${\bar K},N$) spectra for the formation of kaonic nuclei using the absorptive potential obtained in Sec.~\ref{sec:3}.
We also show the possible background contributions to the emitted proton spectrum.
Sec. \ref{sec:5} is devoted to conclusions of this paper.

\section{\label{sec:2}Effective Nuclear density probed by Kaonic Atoms}
In this section, we would like to clarify the meanings of the experimental information concerning the kaon absorption. We find the experimental information of the absorption processes in kaon--nucleus bound systems in Ref.~\cite{vander77}, where the branching fractions for $K^-$ absorption in $^{12}$C were determined experimentally.
The branching ratios of one--body mesic decay (${\bar K}N\rightarrow \pi Y$) and two--body non--mesic decay (${\bar K}NN\rightarrow YN$) were concluded to be $80\%$ and $20\%$, which were used to evaluate the widths of the kaonic nuclear states in Refs.~\cite{gata06,mares05}.
This ratio, however, was determined by the stopped kaon absorption from a certain atomic orbit and should be considered as the ratio at the nuclear surface.

\begin{figure}
\resizebox{0.5\textwidth}{!}{%
  \includegraphics{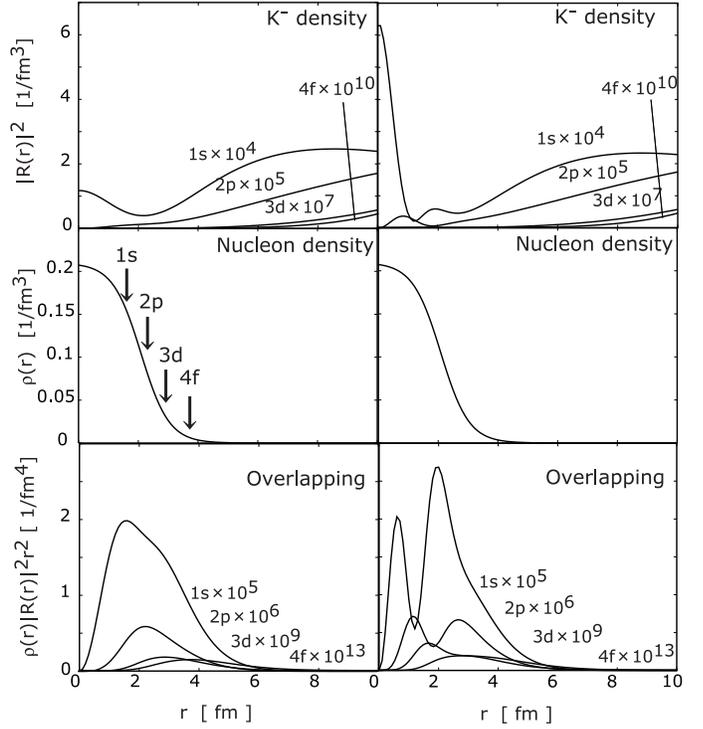}
}
\caption{Overlapping probabilities (lower frame) of the $K^-$ densities (upper frame) with the nucleon densities (middle frame) in the atomic kaon bound states in $^{12}$C calculated with the (left) chiral unitary and (right) phenomenological optical potentials.
The radius $r$, where the overlapping probability take highest value for each atomic state, is shown in middle frame by an arrow  for the chiral unitary potential which indicates the effective nuclear density probed by the kaon in the atomic state.
}
\label{fig:overlap}       
\end{figure}

First, we examine which part ($r$) of the nuclear density ($\rho(r)$) is probed by a kaon in the atomic state using a similar analysis as in Ref.~\cite{yamazaki03} for pionic atoms. 
We consider kaonic atoms in $^{12}$C and two kind of optical potentials as kaon--nucleus interaction, which are the chiral unitary potential~\cite{ram} and the phenomenological potential~\cite{batty97}.
The phenomenological potential has much stronger attraction than the chiral unitary potential and has three kaonic nuclear states inside the nucleus.
The proton and neutron matter density distributions are assumed to take two--parameter Fermi distributions with the radius parameter $R=2.0005~{\rm fm}$ and the diffuseness parameter $a=2.3/4\ln 3~{\rm fm}$~\cite{atomicdata}. We take into account the size of the nucleon using the same prescription as in Ref.~\cite{nieves93}. In Fig.~\ref{fig:overlap}, we show the kaonic atom densities ($|R_{\rm nl}(r)|^2$), the nuclear densities ($\rho(r)$) and the overlapping densities (namely, the nuclear densities probed by $K^-$), defined as
\begin{equation}
S(r)=\rho(r)|R_{\rm nl}(r)|^2r^2~,
\label{eq:s}
\end{equation}
for $^{12}$C with the chiral unitary and the phenomenological optical potentials.

From these figures, we find that the peak positions of the overlapping densities, and thus, the effective nuclear densities $\rho_{\rm e}$ probed by a kaon, which are defined as the nuclear density at radius $r$ where the overlapping density takes maximum value, depend on the atomic orbits of bound kaon significantly.
Actually, we find that the kaon in atomic $1s$ state effectively probes $\rho_{\rm e}^{1s}=0.155~{\rm fm}^{-3}$, while for the kaon in atomic $4f$ state the effective density is $\rho_{\rm e}^{4f}=0.00657~{\rm fm}^{-3}$ for the chiral unitary optical potential.
For the phenomenological optical potential, we also find a significant dependence of the peak positions of the overlapping densities to the kaon atomic states.
In this case, the overlapping densities have 2 peaks for kaonic $1s$ and $2p$ states because of the behavior of kaonic densities which indicate the existence of the kaonic nuclear states.
We show the effective nuclear densities in Table~\ref{tab:effden} only for the chiral unitary potential because of the complicated structure of $S(r)$ for the phenomenological potential with double peaks.
We will show below the quantitative results for the phenomenological potential case in a different way.
We should notice here that this strong dependence of the peak positions of the overlapping densities to the atomic states of the kaon is completely different from that of pionic atoms, where the overlapping densities are peaked at a radius slightly less than the half--density radius, nearly independent of the nucleus and the $\pi^-$ quantum numbers~\cite{yamazaki03}. 
This means that the bound $\pi^-$ always probes effectively a fraction of the full nuclear density $\rho_{\rm e}\approx 0.60\rho(0)~$, while the bound $K^-$ probes various nuclear densities according to the atomic states where the $K^-$ is trapped. 
\begin{table}
\caption{The effective nuclear matter densities $\rho_{\rm e}$ of kaonic atoms in $^{12}$C, which are defined as the nuclear matter densities at radius $r$ where the overlapping densities shown in Fig.~\ref{fig:overlap} takes the maximum value.
The density distributions of kaonic atoms are calculated with the chiral unitary potential.}
\label{tab:effden}       
\begin{center}
\begin{tabular}{|c|c|}
\hline
\multicolumn{2}{|c|}{}\\
\multicolumn{2}{|c|}{Chiral Unitary potential}\\
\hline
Atomic States&$\rho_{\rm e}$~fm$^{-3}$\\
\hline
$1s$&0.155\\
\hline
$2p$&8.16$\times~10^{-2}$\\
\hline
$3d$&3.16$\times~10^{-2}$\\
\hline
$4f$&6.57$\times~10^{-3}$\\
\hline
\end{tabular}
\end{center}
\end{table}

 To see this kaonic orbit dependence of $\rho_{\rm e}$ in a different way and to show the results for both potentials in the same manner, we show in Table~\ref{rms} the calculated rms radii $R_{\rm rms}^{\rm ov}$ of overlapping densities $S(r)$ which are defined as,
\begin{equation}
R_{\rm rms}^{\rm ov}=\Bigl[ \frac{\int S(r)r^2 d^3r}{\int S(r)d^3r}\Bigr]^{1/2}~~.
\label{eq:ov}
\end{equation}
We find that the $R_{\rm rms}^{\rm ov}$ depends on the kaonic states significantly both for the chiral unitary and phenomenological potential cases.
The $R_{\rm rms}^{\rm ov}$ with the chiral unitary potential is slightly larger than that with the phenomenological potential.
\begin{table}
\caption{calculated root--mean--square radii of the overlapping densities for four kaonic atom states with the chiral unitary and the phenomenological potentials. See text for the definition of $R_{\rm rms}^{\rm ov}$.}
\label{rms}       
\begin{center}
\begin{tabular}{|c|cccc|}
\hline
&&&&\\
$R_{\rm rms}^{\rm ov}${\rm [fm]}&$1s$&$2p$&$3d$&$4f$\\
&\multicolumn{4}{c|}{}\\
\hline
Chiral Unitary&~~3.60~~&~~3.73~~&~~4.42~~&~~5.31~~\\
Phenomenology&~~3.53~~&~~3.70~~&~~4.18~~&~~5.06~~\\
\hline
\end{tabular}
\end{center}
\end{table}

Actually in the history of kaonic atom studies, it has been recognized for a long time that the kaon in atomic states probed the nuclear surface~\cite{kim71}.
Here we would like to emphasize again that this feature is much different from that of pionic atoms and the nuclear densities probed by kaons depend significantly on the quantum numbers of the atomic states of the kaon.

Investigating the correlations of the potential parameters is another interesting issue.
As indicated by Seki and Masutani, there are some correlations between potential parameters for the cases of shallow pionic atoms~\cite{seki83}, which also hold for deeply bound pionic atoms~\cite{toki89}.
These correlations are known to indicate that the pionic atom properties are determined by the potential strength at a certain nuclear matter density, which is $\rho \sim \displaystyle\frac{1}{2}\rho_0$ for the local part of pion--nucleus optical potential for all atomic states.
This value is close to the $\rho_{\rm e}$ of pionic atoms determined by using the overlapping density $S(r)$~\cite{yamazaki03}.

\begin{figure}
\resizebox{0.5\textwidth}{!}{%
  \includegraphics{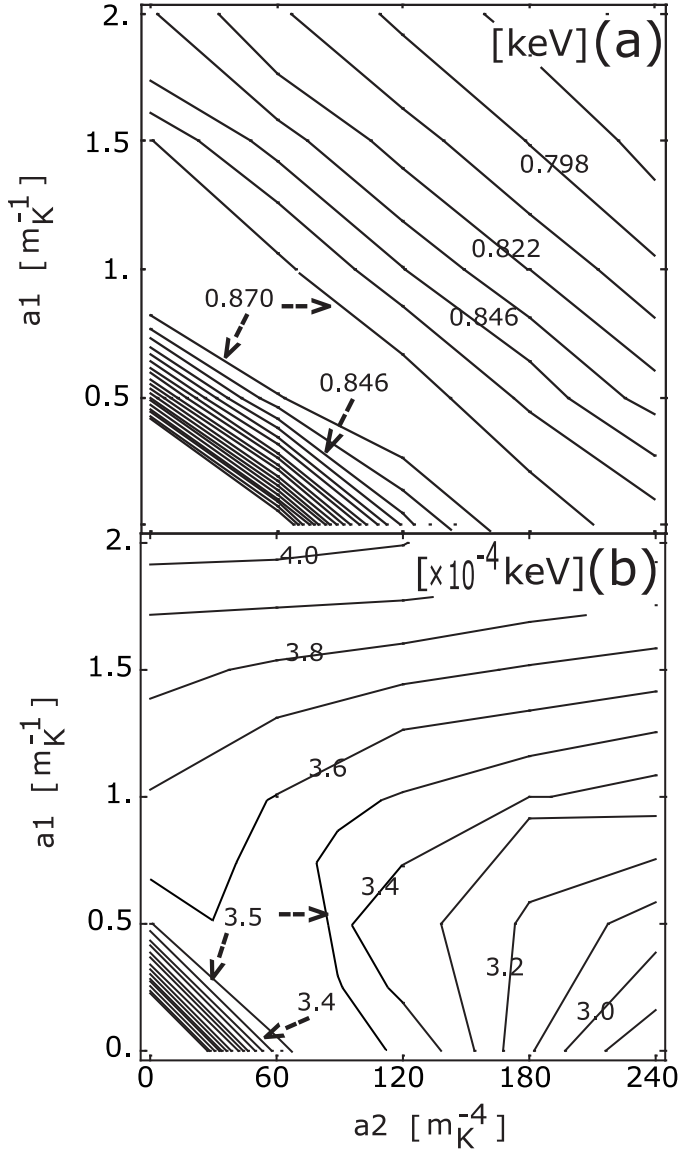}
}
\caption{Contour plots of the widths for the (a) $2p$ and (b) $3d$ states of kaonic atoms in $^{12}$C in the $a_1-a_2$ plane of the imaginary part of the optical potential defined in Eq.~(\ref{eq:3}).
The real part of the optical potential is fixed to be the same as the phenomenological optical potential~\cite{batty97}.
}
\label{fig:2pcontour}       
\end{figure}

In the present case for kaonic atoms, the effective nuclear matter densities $\rho_{\rm e}$ probed by kaonic atoms are significantly different for different atomic states.
Thus, we can expect to have different correlations between potential parameters for different atomic states, which could be used to determine the potential parameters using the atomic data.
Since we are interested in the absorption processes of the kaon in the nucleus, we study the correlation of the parameters of the imaginary part of the optical potential defined in the next section in Eq.~(\ref{eq:3}).
The real part of the optical potential is fixed to be the same as for the phenomenological potential~\cite{batty97}.
We show the contour plots of the calculated widths of kaonic $^{12}$C atoms in $2p$ and $3d$ states in Fig.~\ref{fig:2pcontour}.
As we expected, we find much different parameter correlations for these states.
We mention here again that these different features of the contour plots are expected to be due to the strong atomic state dependence of the nuclear matter densities probed by the kaonic atoms.
This feature of kaonic atoms is completely different from that of pionic atoms.

\section{\label{sec:3}One--body and two--body absorption strength}
Now, we need to reexamine the meanings of the experimental information on the kaon absorption obtained in Ref.~\cite{vander77}, where the branching ratio of one--body and two--body $K^-$ absorption processes are determined to be 80 $\%$ and 20 $\%$.
Since, as discussed in Sec.~\ref{sec:2} in detail, the effective nuclear matter densities probed by the kaonic atoms depend significantly on the atomic states where kaon exists, it is quite important to know the atomic state from where most of the kaons are absorbed by the nucleus and to know the nuclear matter density probed by the deexcitation process when emitting the X--ray.

For this purpose, we assume the imaginary part of the optical potential to take the form,
\begin{equation}
{\rm Im}V_{\rm opt}=-\frac{4\pi}{2\mu} \Bigl{\{}\bigl( 1+\frac{m_{\bar K}}{m_N}\bigr)a_1\rho+\bigl(1+\frac{m_{\bar K}}{2m_N}\bigr)a_2\rho^2\Bigr{\}}
\label{eq:3}
\end{equation}
to treat the one--body and two--body processes separately in a clear manner and consider the parameters $a_1$ and $a_2$.
In Ref.~\cite{mares05}, we can find another kind of analysis considering different forms of the imaginary potential.

We then consider the experimental information of the kaon absorption~\cite{vander77}.
The nuclear matter density $\rho_{\rm abs}$ probed by the kaon absorption experiment is expected to satisfy the condition
\begin{equation}
\bigl(1+\frac{m_{\bar K}}{m_N}\bigr)a_1\rho_{\rm abs}:\bigl(1+\frac{m_{\bar K}}{2m_N}\bigr)a_2\rho_{\rm abs}^2=80:20~~,
\label{eq:rho_new}
\end{equation}
because of the consequence of the experiment~\cite{vander77}, and this means the existence of the linear relation between $a_1$ and $a_2$ as,
\begin{equation}
a_1=\frac{80}{20}\frac{(1+\frac{m_{\bar K}}{2m_N})}{(1+\frac{m_{\bar K}}{m_N})}\rho_{\rm abs}a_2~~.
\label{eq:5}
\end{equation}
By assuming the $\rho_{\rm abs}$ in Eq.~(\ref{eq:5}), we can determine the linear relation between the coefficients $a_1$ and $a_2$. One of the reasonable assumption is to fix the $\rho_{\rm abs}$ to be the same value as the $\rho_{\rm e}$ of the atomic states considered in Sec.~\ref{sec:2}.

To know the atomic state where the kaon is mainly absorbed in the nucleus in the deexcitation process, we can expect to have the experimental information from the data of the absolute intensities of the emitted X--ray shown in Table~${\rm I}\hspace{-.1em}{\rm V}$ of Ref.~\cite{clyde74} for the kaonic atoms in $^{12}$C.
In that Table, we can find that the observed X--ray intensities from transitions $(n_{\rm i},n_{\rm f})=(3,2)$ are around one order of magnitude smaller than those from the transitions $(n_{\rm i},n_{\rm f})=(4,3)$, where $n_{\rm i}$ and $n_{\rm f}$ are the principal quantum numbers of the initial and final kaonic states.
Thus, we may expect naively that most of the kaons are absorbed by the nucleus from the $3d$ atomic state and a tiny fraction of the kaon continue the X--ray transition to deeper $2p$ state.
We consider both possibilities of kaon absorption from atomic $2p$ and $3d$ states, here.

Based on these facts, we postulate here the following in order to determine the potential parameters of the imaginary part in a phenomenological way;

(i) most of the kaon is absorbed by the nucleus from $2p$ and $3d$ atomic states in the kaon--$^{12}$C system in the X--ray deexcitation processes, and the $\rho_{\rm abs}$ in Eq.~(\ref{eq:5}) is equal to be $\rho_{\rm e}$ in Table~\ref{tab:effden},

(ii) the depth of the imaginary potential is the same as that of the phenomenological potential obtained in Ref.~\cite{batty97} to be roughly consistent with the accumulated atomic data, and additionally we take the real part of the same phenomenological potential.

By these assumptions, we can fix the potential parameters as;
\begin{eqnarray}
&{\rm Set1:}& (a_1,a_2)=(0.883, 51.2) {\rm ~for~} \rho_{\rm abs}=\rho_{\rm e}(2p)~,~ \nonumber\\
&{\rm Set2:}&(a_1,a_2)=(0.547, 81.9){\rm ~for~}\rho_{\rm abs}=\rho_{\rm e}(3d)~,~
\label{eq:a1a2}
\end{eqnarray}
in units of $a_1$ [$m_K^{-1}$] and $a_2$ [$m_K^{-4}$], respectively.

\begin{figure}
\resizebox{0.5\textwidth}{!}{%
  \includegraphics{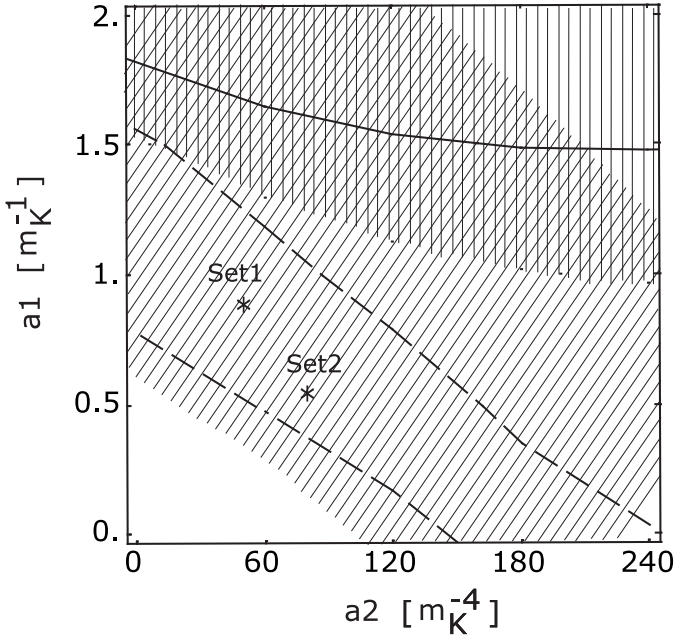}
}
\caption{Experimental data of the width of kaonic $2p$ state $\Gamma_{2p}$ (dashed lines) and the X--ray energy of the $3d\rightarrow 2p$ transition (solid line) are shown in the plane of the potential parameters $a_1$ and $a_2$ for the imaginary potential Eq.~(\ref{eq:3}) for $^{12}$C.
The real part of the optical potential is fixed to be the same as the phenomenological optical potential~\cite{batty97}.
The errors of both data are indicated by the hatched areas.
Please note there are two dashed lines for one value of the experimental width because of the structure of the contour of the $\Gamma_{2p}$ as shown in Fig.~\ref{fig:2pcontour} (a).
Two stars in the figure indicate the parameter sets ($a_1,a_2$) determined in Eq.~(\ref{eq:a1a2}).
See text for details.
}
\label{fig:expcontour}       
\end{figure}

We show in Fig.~\ref{fig:expcontour} the region of $a_1$ and $a_2$ parameters consistent with the experimental data~\cite{back72} of the width of the kaonic $2p$ state and the $3d\rightarrow 2p$ transition X--ray energy in $^{12}$C.
Since the contour for the width of the $2p$ state has the ridge structure, there are two dashed lines for one value of the experimental width.
We find that the errors of the data are too large to determine the unique parameter set $(a_1,a_2)$ from the data.
However, we can find the acceptable region of the parameters $a_1$ and $a_2$.
We show the values of the ($a_1,a_2$) parameters determined above in Eq.~(\ref{eq:a1a2}) in the same figure, and we find that both sets of parameters are close to the acceptable region determined from the data.
We can fix the parameter sets ($a_1, a_2$) so as to reproduce both experimental data (energy shift and width) instead of imposing the condition (ii) described above.
We, however, adopted the condition (ii) so as to avoid obtaining a significantly deeper ${\rm Im}V_{\rm opt}$ than that of the phenomenological potential.

\begin{figure*}
\resizebox{1.0\textwidth}{!}{%
  \includegraphics{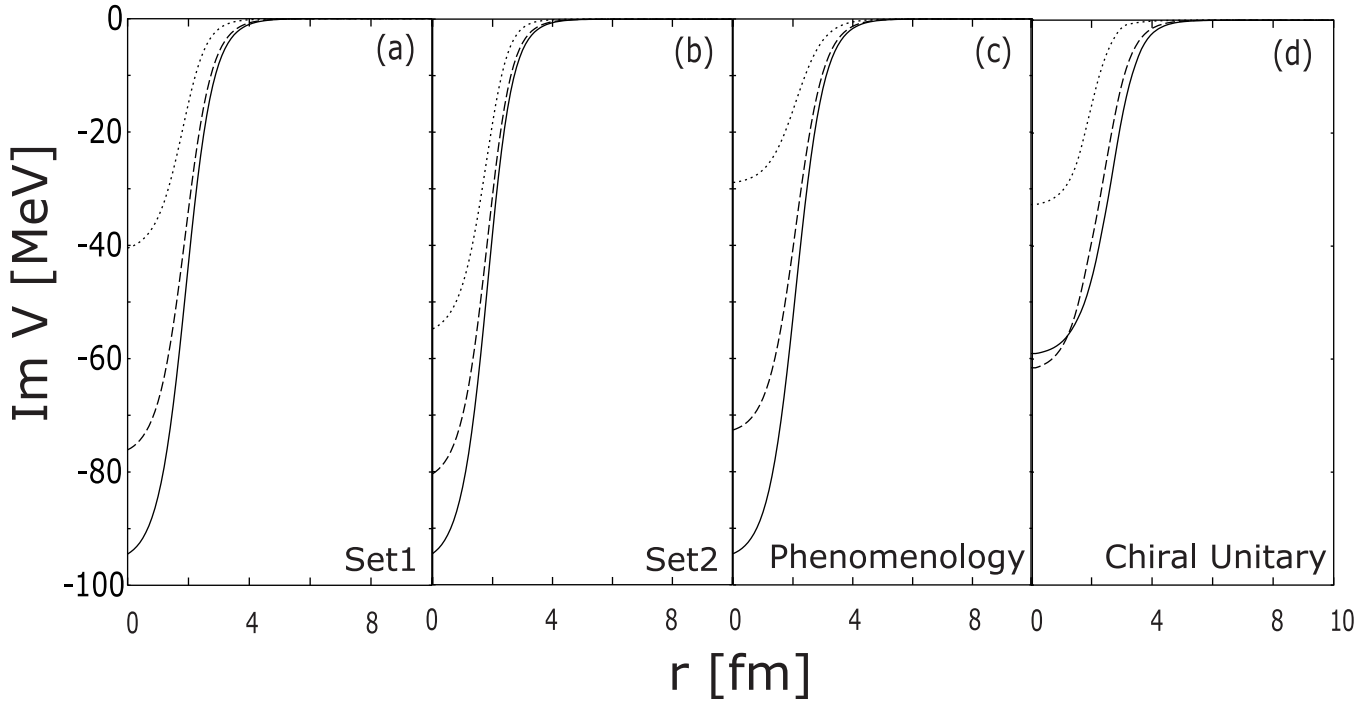}
}
\caption{The imaginary part of the kaon--nucleus optical potentials for $^{12}$C are plotted as functions of the radial coordinate $r$ for three kaon  energies $E=0$ (solid line) , $-50$ (dashed line), and $-100$ (dotted line) MeV.
The potentials shown in (a) and (b) are obtained in Eq.~(\ref{eq:a1a2}) (please see text for details).
The phenomenological potential in Ref.~\cite{batty97} with the energy dependence  adopted in Refs.~\cite{gata06,mares05} is shown in (c).
The imaginary potential of the chiral unitary model \cite{ram} is shown in (d).
}
\label{fig:ImVopt}       
\end{figure*}

We show the imaginary part of the optical potential in Fig.~\ref{fig:ImVopt} for different kaon energies.  The energy dependence is introduced for phenomenological potentials as, 
\begin{eqnarray}
{\rm Im}V_{\rm opt}(r,E)&=&-\frac{4\pi}{2\mu}\bigl\{\bigl(1+\frac{m_{\bar K}}{m_N}\bigr)a_1\rho(r)f_1(E) \nonumber \\
&&+\bigl(1+\frac{m_{\bar K}}{2m_N}\bigr)a_2\rho^2(r)f_2(E)\bigr\}
\label{eq:ImV6}
\end{eqnarray}
using the phase space factors $f_1$ and $f_2$ defined in Ref.~\cite{mares05} and used in Ref.~\cite{gata06} for one--body and two--body absorption processes, respectively. We also show in the same figure the optical potentials of the phenomenological model obtained in Ref.~\cite{batty97} with the same energy dependence  adopted in Refs.~\cite{gata06,mares05}, and the theoretical optical potential obtained by the chiral unitary model in Ref.~\cite{ram} for comparison.
As we can see from the figures, the three imaginary potentials (a-c) which have the same depth at the nuclear center at the threshold energy have significantly different energy dependence.
In Fig.~\ref{fig:ImVopt}(a), the observed ratio of the one--body (80$\%$) and two--body (20$\%$) absorption~\cite{vander77} is interpreted as the ratio at nuclear density $\rho=8.16\times 10^{-2}$ fm$^{-3}$ probed by kaon in atomic $2p$ state as described in Sec.~\ref{sec:2}.
Similarly, in Fig.~\ref{fig:ImVopt}(b), the observed ratio is interpreted to be the ratio at nuclear density probed by kaon in atomic $3d$ state $\rho=3.16\times 10^{-2}$ fm$^{-3}$.
In these cases, the ratio of the one--and two--body processes at the nuclear center is different from that determined by the experiment which is expected to provide the information at nuclear surface.
And, due to the different energy dependences of both processes as expressed by the functions $f_1$ and $f_2$, the energy dependence of the whole imaginary potentials is different for these cases.
In Fig.~\ref{fig:ImVopt}(c), the one--body to two--body ratio is assumed to be the same for all nuclear density region for obtaining the energy dependence as in Ref.~\cite{gata06}, and, hence, the energy dependence of the imaginary potential is different from both (a) and (b).
The imaginary potential of the chiral unitary model is also shown in Fig.~\ref{fig:ImVopt}(d) for comparison.

\section{\label{sec:4}Formation spectra of kaonic nuclei by In-flight ($K^-,p$) reactions}
\begin{figure}
\resizebox{0.5\textwidth}{!}{%
  \includegraphics{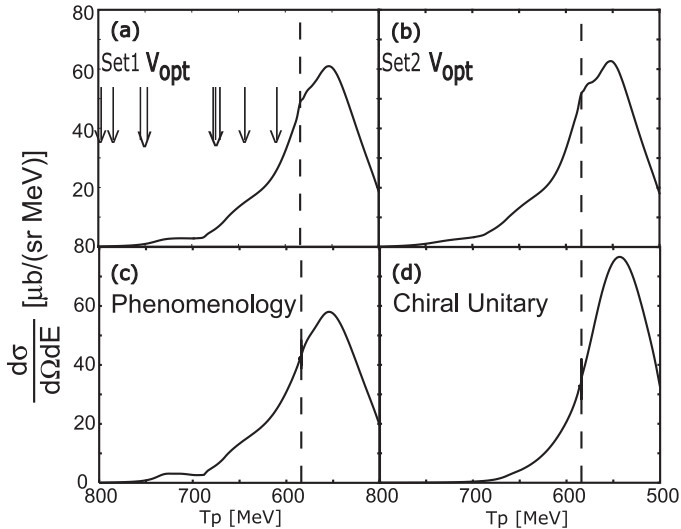}
}
\caption{Calculated spectra of the $^{12}$C($K^-,p$) reactions for the formation of kaon--nucleus systems at $T_{K^-} = 600$ MeV plotted as functions of the emitted proton energy $T_p$ at $\theta_p^{\rm Lab} = 0$ (deg.) with the energy dependent optical potentials.
The imaginary part of the optical potentials for (a) and (b) are obtained in Eq.~(\ref{eq:a1a2}) (please see text for details).
The spectra in (c) and (d) are calculated with the phenomenological optical potential~\cite{batty97} and the chiral unitary potential~\cite{ram} same as those reported in Ref.~\cite{gata06}.
The vertical dashed lines indicate the kaon emission threshold in the final states.
The arrows indicate the proton energies from possible background processes listed in Table \ref{bg} estimated without final state interactions.
}
\label{fig:12C}       
\end{figure}

We have shown the calculated ($K^-,p$) spectra for the formation of kaonic nuclear states in Fig.~\ref{fig:12C} using the same theoretical reaction models described in Ref.~\cite{gata06}.
We have found that the three potentials, which have the same real part and almost the same depth of the imaginary part at $E=0$, provide a slightly different spectra in energy regions of the kaonic nuclear states.
As we can  see in Fig.~\ref{fig:12C} (c), the spectrum obtained by the phenomenological optical potential~\cite{batty97} with the assumed energy dependence for the imaginary part shows some indications of kaonic nucleus formation at $T_p= 700- 750$MeV as reported in Ref.~\cite{gata06}.
On the other hand if we assume that the kaon absorption reported in Ref.~\cite{vander77} occurs at the nuclear surface probed by the kaonic $2p$ and $3d$ atomic sates, we find that smoother spectra are expected as shown in Fig.~\ref{fig:12C} (a) and (b) due to the different energy dependence of the imaginary part of the optical potential.
And we think that it is more difficult to observe the signals of the kaonic nuclei formation in these cases as can be seen in Fig.~\ref{fig:12C} (a) and (b) around  $T_p= 700- 750$MeV.
All our calculated results reported here and in Ref.~\cite{gata06} show no clear peak structures for the formation of kaonic nuclear states in $^{12}$C and indicate the difficulties to obtain reliable evidence of the existence of kaonic nuclear states.
Our theoretical results seems fully consistent with the latest data by Kishimoto et al.~\cite{kishimoto06}.

We also make some comments on the background processes in this ($K^-,p$) reactions.  As pointed out in Refs.~\cite{oset06,magas06}, some familiar quasi--free processes may have contributions to the ($K^-,p$) spectra. 
\begin{table}
\caption{Possible quasi--free background contributions to in--flight ($K^-,p$) reactions for the formation of kaon--nucleus bound systems, which are classified as proton emissions (indicated by circles) from two body kaon absorption (1,2), hyperon decay after two body kaon absorption (3,4), hyperon decay after one body kaon absorption (5,6), two body pion absorption after one body kaon absorption (7-10).
The emitted proton energies listed in the last column are evaluated by assuming the nucleons at rest with the 8 MeV binding in the nucleus. 
The background contributions are expected to appear in the spectra at lower proton energies than those listed in this table because of the final state interactions, which are not considered here.
}
\label{bg}       
\begin{center}
\begin{tabular}{|c|l|l|c|}
\hline
&\multicolumn{2}{c|}{~}&\\
~&\multicolumn{2}{c|}{Process}&$T_p$\\
&\multicolumn{2}{c|}{~}&[MeV]\\
\hline
~~1~~~&~$K^-pp\rightarrow \Lambda~p\hspace{-.75em}\bigcirc$~&&~798.89~~\\
~~2~~~&~$K^-NN\rightarrow \Sigma~p\hspace{-.75em}\bigcirc$~&~&~749.32~~\\
~~3~~~&~$K^-NN\rightarrow N\Sigma$~&~$\Sigma\rightarrow \pi~p\hspace{-.75em}\bigcirc$~&~816.59~~\\
~~4~~~&~$K^-pN\rightarrow N\Lambda$~&~$\Lambda\rightarrow \pi^-~p\hspace{-.75em}\bigcirc$~&~785.94~~\\
~~5~~~&~$K^-p\rightarrow \pi^-\Sigma^+$~&~$\Sigma^+\rightarrow \pi^0~p\hspace{-.75em}\bigcirc$~&~644.71~~\\
~~6~~~&~$K^-p\rightarrow\pi^0\Lambda$~&~$\Lambda\rightarrow \pi^-~p\hspace{-.75em}\bigcirc$~&~610.35~~\\
~~7~~~&~$K^-p\rightarrow \pi^-\Sigma^+$~&~$\pi^-pp\rightarrow n ~p\hspace{-.75em}\bigcirc$~&~678.67~~\\
~~8~~~&~$K^-p\rightarrow\pi^0\Sigma^0$~&~$\pi^0pN\rightarrow N~p\hspace{-.75em}\bigcirc$~&~676.16~~\\
~~9~~~&~$K^-p\rightarrow \pi^+\Sigma^-$~&~$\pi^+nN\rightarrow N~p\hspace{-.75em}\bigcirc$~&~671.47~~\\
~~10~~~&~$K^-p\rightarrow \pi^0\Lambda$~&~$\pi^0 p N\rightarrow N~p\hspace{-.75em}\bigcirc$~&~756.78~~\\
\hline
\end{tabular}
\end{center}
\end{table}
We have checked more than 100 combinations of the possible processes which provide the proton emission in the final states after kaon absorption.
We found 10 processes listed in Table~\ref{bg} which will provide the final proton in the same energy region for the kaonic nucleus formation.
In our evaluation of the proton kinematics, we have assumed that the initial nucleons are at rest in the nucleus with the 8 MeV binding, and we have neglected the final states interaction effects such as multiple scatterings.
Thus, we expect that the real contributions of the background distribute for lower proton energy regions than the proton energies listed in Table~\ref{bg}.
We also show the proton energies in the Fig.~\ref{fig:12C} (a) by the solid arrows.
In addition, it might be important to consider the proton emissions by the quasi--elastic reactions with target break--up as the background.
We think that to know the spectra of kaonic nucleus formation we should be very careful to remove these background contributions from observed spectra by using the reliable theoretical calculations.

\section{\label{sec:5}Conclusions}
Kaonic atoms and kaonic nuclei are very important objects to extract valuable information on the kaon behaviors at finite densities.
In addition, the possible existence of kaonic nuclear states with narrow widths and the exotic states with higher densities predicted theoretically stimulated both theorists and experimentalists, leading to active research concerning the structure and formation of the kaon--nucleus systems, recently.
In order to make these active researches very fruitful as they should be, we believe that we need definitely the theoretical calculations for the formation reaction spectra to get decisive results from the experimental data.

In this paper, we have considered the kaon absorption processes from the atomic orbits into nucleus and reexamined the meaning of the experimental data, which determined the ratio of the one--body and two--body absorptive strengths by the stopped $K$ experiment.
We find that the effective nuclear density probed by kaons in the atomic state depends strongly on the atomic orbit of kaon.
We estimate that the nuclear density probed by kaons in the atomic $1s$ state is roughly 20 times larger than that by the atomic $4f$ state (see Table~\ref{tab:effden}).
This feature is much different from that of pionic atoms in which the nuclear density probed by pions is nearly independent both on the atomic orbits and the nuclide.
Thus, we need to know the atomic orbit where kaon absorption occurs to interpret the meaning of the absorption ratio of the one--body and two--body processes.
We have considered two cases by assuming that the kaon absorption mainly occurs at (i) atomic $2p$ state and (ii) atomic $3d$ state.
By considering the phenomenological optical potentials, we show the differences of the energy dependence of the imaginary part of the optical potentials for these two cases.
As a natural consequence, if the atomic kaon probes the smaller nuclear density, the ratio of the two--body absorption at nuclear center is larger than the observed value, and the depth of the imaginary potential is deeper even at smaller kaon energies as in kaonic nuclear states because of the large phase space for the two--body processes.

We also investigate the effects to the $(K^-,p)$ spectra for the formation of the kaonic nuclear states.
We find that the reaction spectra are even more smooth than those reported in Ref.~\cite{gata06} and the signals become more unclear if the absorption happens at the nuclear surface.
We conclude again that to observe the peak structure for the evidence of kaonic nuclear state is extremely difficult for $K^--^{12}$C systems.

We also consider the possible quasi--free kaon absorption processes which will be the background of the formation spectra by ($K^-,p$) reaction (Table~\ref{bg}).
These background processes and the quasi--elastic reactions with target break--up seem to be important to extract the signals from the data.\\

We would like to express our sincere thanks to H. Nagahiro for many discussions and collaborations on this subject.
We would like to thank many useful discussions and comments for E. Oset, H. Toki and A. Hosaka.
We would like to thank D. Jido for stimulating discussions on meson properties at finite density.
We also thank to A. Gal for useful discussions.
This work is partly supported by Grants-in-Aid for scientific research of JSPS  
(No. 16540254) and by CSIC and JSPS under the Japan-Spain Research Cooperative Program.
The authors thank the Yukawa Institute for Theoretical Physics at Kyoto University.
Discussions during the YKIS2006 on "New Frontiers QCD" were useful to complete this work.

%

\begin{thebibliography}{99}
\bibitem{Kishimoto99} T. Kishimoto, Phys. Rev. Lett. {\bf 83}, 4701 (1999).
\bibitem{Kishimoto03} T. Kishimoto et al., Prog.~Theor.~Phys.~Suppl.~No.{\bf 149}, 264 (2003), T. Kishimoto et al., Nucl. Phys. {\bf A754}, 383c (2005). 
\bibitem{gata05} J. Yamagata, H. Nagahiro, Y. Okumura and S. Hirenzaki, Prog. Theor. Phys. 114, 301 (2005).~(Errata 114, 905 (2005).)
\bibitem{gata06} J. Yamagata, H. Nagahiro and S. Hirenzaki, Phys. Rev. C {\bf 74}, 014604 (2006).
\bibitem{finuda} M. Agnello et al., Phys. Rev. Lett. {\bf 94}, 212303 (2005).
\bibitem{Akaishi02} Y. Akaishi and T. Yamazaki, Phys. Rev. C {\bf 65}, 044005 (2002).
\bibitem{dote04} A. Dot$\acute{\rm e}$, H. Horiuchi, Y. Akaishi and T. Yamazaki, Phys. Lett. B {\bf 590}, 51 (2004).
\bibitem{oset06}E. Oset and H. Toki, Phys. Rev. C {\bf 74}, 015207 (2006).
\bibitem{magas06} V. K. Magas, E. Oset, A. Ramos, and H. Toki, Phys. Rev. C {\bf 74}, 025206 (2006). 
\bibitem{finuda06}M. Agnello et al., Nucl. Phys. {\bf A775}, 35 (2006).
\bibitem{Iwasaki03} M. Iwasaki et al., nucl-ex/0310018.\\
T. Suzuki et al., Phys. Lett. B {\bf 597}, 263 (2004); Nucl. Phys. {\bf A754} 375c (2005).
\bibitem{iwasaki06}M. Iwasaki, talks at Japan Physical Society Meeting,
Nara, Japan, Sep. 2006 and at YKIS 2006, Kyoto, Japan, Dec. 2006; H. Outa, talk at Korean Physical Society Meeting, Daegu, Korea, Oct. 2006.

\bibitem{batty97} C. J. Batty, E. Friedman and A. Gal, Phys.~Rep. {\bf 287}, 385 (1997). 
\bibitem{waa} T. Waas, N. Kaiser and W. Weise, Phys. Lett. B {\bf365}, 12 (1996); {\bf 379}, 34 (1996); T. Waas and W. Weise, Nucl. Phys. {\bf A625}, 287 (1997).
\bibitem{lut} M. Lutz, Phys. Lett. B {\bf 426}, 12 (1998).
\bibitem{ram} A. Ramos and E. Oset, Nucl. Phys. {\bf A671}, 481 (2000).
\bibitem{ciep01} A. Ciepl$\acute{\rm y}$, E. Friedman, A. Gal and J. Mare$\check {\rm s}$, Nucl. Phys. {\bf A696}, 173 (2001).
\bibitem{tolo01} L. Tol$\acute{\rm o}$s, A. Ramos, A. Polls, T. T. S. Kuo, Nucl. Phys. {\bf A690}, 547 (2001).
\bibitem{tolo02} L. Tol$\acute{\rm o}$s, A. Ramos, A. Polls, Phys. Rev. C {\bf 65}, 054907 (2002).
\bibitem{schaffner00}J. Schaffner--Bielich, V. Koch, M. Effenberger, Nucl. Phys. {\bf A669}, 153 (2000).
\bibitem{hirenzaki00} S. Hirenzaki, Y. Okumura, H. Toki, E. Oset and A. 
Ramos, Phys. Rev. C {\bf 61}, 055205 (2000).
\bibitem{baca00}A. Baca, C. Garc$\acute{\rm i}$a--Recio, J. Nieves, Nucl. Phys. {\bf A673}, 335 (2000). 
\bibitem{friedman99} E. Friedman and A. Gal, Phys. Lett. B {\bf 459}, 43 (1999).
\bibitem{friedman99_2} E. Friedman and A. Gal, Nucl. Phys. {\bf A658}, 345 (1999). 
\bibitem{vander77}C. Vander Velde--Wilquet, J. Sacton, J. H. Wickens, D. N. Tovee, D. H. Davis, Nuovo Cimento A 39, 538 (1977).
\bibitem{mares05} J. Mare$\check{\rm s}$, E. Friedman and A. Gal, Phys. Lett. B {\bf 606}, 295 (2005); Nucl. Phys. {\bf A770}, 84 (2006). 
\bibitem{yamazaki03}T. Yamazaki, S. Hirenzaki, Phys. Lett. B {\bf 557}, 20 (2003). 
\bibitem{atomicdata}G. Fricke, et al., ATOMIC DATA AND NUCLEAR DATA TABLES {\bf 60}, 177 (1995).
\bibitem{nieves93}J. Nieves, E. Oset, and C. Garcia--Recio, Nucle. Phys. {\bf A554}, 509 (1993).
\bibitem{kim71}For example, see 'Mesic atoms and nuclear structure', Y. N. Kim, North--Holland Pub., (1971).
\bibitem{seki83}R. Seki, K. Masutani, Phys. Rev. C 27, 2799 (1983).
\bibitem{toki89}H. Toki and T. Yamazaki, Phys. Lett. B {\bf 213}, 129 (1988), H. Toki, S. Hirenzaki, R. S. Hayano, T. Yamazaki, Nucl. Phys. {\bf A 501}, 653 (1989), 
S. Hirenzaki, H. Toki and T. Yamazaki, Phys. Rev. C {\bf 44}, 2472 (1991),
H. Toki, S. Hirenzaki and T. Yamazaki, Nucl. Phys. {\bf A530}, 679 (1991).
\bibitem{clyde74}Clyde E. Wiegand and Gary L. Godfrey, Phys. Rev. A {\bf 9}, 2282 (1974).
\bibitem{back72}G. Backenstoss et al., Phys. Lett. B {\bf 38}, 181 (1972). 
\bibitem{kishimoto06}T. Kishimoto, talks at Korean Physical Society Meeting, Daegu, Korea, Oct. 2006 and at YKIS 2006, Kyoto, Japan, Dec. 2006.

\end{thebibliography}
%

\end{document}